\documentclass{QIC}
\usepackage{upgreek}

\allowdisplaybreaks
\def\crauthorname{Mobin Motaharifar et~al.}
\begin{document}
\headtitle{Quantum Information \& Computation\\
ISSN 1533-7146 Volume 25 (2025) pp. 175--194}

%\doiurl{https://doi.org/xxxxxxxx}
\doiurl{DOI: 10.2478/qic-2025-0009}
	
\title{A Survey on Continuous Variable Quantum Key Distribution for Secure Data Transmission: Toward the Future of Secured Quantum-Networks\vspace*{-5pt}}

\author{Mobin Motaharifar\affilmark{~}*${}^{\dagger}$, Mahmood Hasani\affilmark{~}${}^{\dagger}$ and Hassan Kaatuzian\affilmark{~}*}

\correspond{Correspondence author: mobinmotaharifar@aut.ac.ir (M.M.); hsnkato@aut.ac.ir (H.K.)\\
${}^\dagger$ \ Both authors have the same contribution to this work.}

\affilbox{~}{Photonics Research Laboratory, Electrical Engineering Dept, Amirkabir University of Technology, Tehran 159163-4311, Iran; spew@aut.ac.ir}

%%%%%  for date information
\received{13 December 2024}
\accepted{9 March 2025}
\published{10 April 2025\vspace*{-6pt}}

\Abstract{Quantum key distribution (QKD) represents a cornerstone of secure communication in the quantum era. While discrete-variable QKD (DV-QKD) protocols were historically the first to demonstrate secure key exchange, continuous-variable QKD (CV-QKD) has emerged as a more practical alternative due to its seamless compatibility with current telecommunications infrastructure. CV-QKD relies on coherent and squeezed states of light, offering significant advantages for integration into modern optical networks. This review comprehensively explores the theoretical underpinnings, technological advancements, and practical challenges of CV-QKD. Special attention is given to the role of photonic integrated circuits (PICs) in enabling scalable and efficient implementation of CV-QKD systems. Furthermore, recent advances in machine learning have been leveraged to optimize CV-QKD performance, with data-driven techniques enhancing noise estimation, parameter optimization, and system security. Additionally, tensor networks provide efficient computational tools for analyzing complex quantum correlations, improving the efficiency and robustness of quantum key distribution protocols. These developments, combined with ongoing improvements in quantum photonic integration, pave the way for the practical deployment of large-scale, high-speed quantum-secure networks.}


\maketitle
%%%%%%%%%%%%%%%%%%%%%%%%%%%%%%%%%%%%%%%%%%%%%%%%%%%%%%%%%%%%%%%%%%%%%%%%%%%%%%% main text start

\vspace*{-4pt}
\section{Introduction}

Quantum communication has emerged as a fundamental pillar of quantum information science, offering unparalleled security through the principles of quantum mechanics. The two primary goals of quantum communication include the advancement of quantum key distribution (QKD) systems and the establishment of distributed quantum computation via a quantum internet. Several large-scale QKD networks have been developed and tested worldwide, including in Switzerland~\cite{1}, Japan~\cite{2}, and the United States~\cite{3}. These networks employ various implementation methods, such as fiber-optic communication, free-space transmission, and satellite-based QKD~\cite{4}. However, the miniaturization and integration of QKD hardware remain an active area of research, as compact and efficient implementations are necessary for practical deployment.

Silicon-based photonic systems have enabled on-chip QKD~\cite{5}, offering numerous advantages. Various materials have been explored for chip-level QKD integration, including indium phosphide (InP)~\cite{6}, potassium titanyl phosphate (KTP)~\cite{7}, and lithium niobate (LiNbO3)~\cite{8}. These materials facilitate the creation of on-chip lasers and high-speed modulators. Silica, known for its low-loss delay lines and fiber compatibility, lacks fast modulation capabilities~\cite{9,10}. Conversely, silicon benefits from mature microfabrication techniques, making it well-suited for photonic component development \cite{11,12,13}.

QKD can be categorized into three main types: discrete-variable quantum key distribution (DV-QKD), continuous-variable quantum key distribution (CV-QKD), and distributed phase-reference coding. DV-QKD relies on the \hbox{quantum} properties of individual photons, such as polarization or phase, while CV-QKD encodes information in the \hbox{continuous} variables of quantum states, such as the quadratures of coherent or squeezed light modes. DV-QKD has been \hbox{studied} extensively and has been successfully demonstrated over transmission distances of approximately 400 km using ultra-low-loss fibers \cite{14,15}. The distributed-phase-reference method, distinct from DV- and CV-QKD, encodes information based on the phase difference between successive pulses or the timing of photon arrivals. Photon counting and post-selection are integral to both DV-QKD and distributed-phase-reference methods, whereas CV-QKD employs homodyne detection for secure key generation~\cite{16}.

The history of DV-QKD dates back to 1984 when Bennett and Brassard introduced the BB84 protocol, which relies on the principle that quantum states must be measured on the correct basis for accurate key generation~\cite{17}. The security of BB84 is underpinned by the no-cloning theorem, which prohibits the exact replication of quantum states. In 1992, Bennett proposed a simplified version, the B92 protocol, which utilizes only two non-orthogonal states for encoding~\cite{18}. Another significant development in DV-QKD was the E91 protocol, introduced by Ekert in 1991, which leverages entangled qubit pairs and employs Bell's test for security verification~\cite{19}.

Continuous Variable Quantum Key Distribution (CV-QKD) utilizes continuous variables of quantum states, typically the quadratures of coherent states or squeezed light, for encoding information. CV-QKD systems are compatible with conventional telecommunication infrastructure, utilizing commercial lasers and homodyne detectors. This compatibility makes CV-QKD particularly appealing for integration with existing communication networks, as it offers high secret key rates over metropolitan-scale distances \cite{20,21,22}. The transition of CV-QKD from laboratory research to real-world implementation is progressing rapidly. These systems provide a promising approach for secure communication, offering advantages such as room-temperature operation and higher key rates compared to DV-QKD. The field continues to evolve, with ongoing research focusing on practical deployment and enhanced security measures. Furthermore, squeezed light plays a crucial role in advancing quantum networks, benefiting both QKD and distributed quantum computing applications \cite{23,24}.

Gaussian quantum information \cite{25,26,27} forms a cornerstone of continuous-variable quantum protocols, leveraging the mathematical elegance and experimental accessibility of Gaussian states. In this framework, quantum states such as coherent and squeezed states are fully characterized by their first and second moments, allowing for efficient description using covariance matrices. Gaussian operations---implemented via beam splitters, phase shifters, and homodyne detectors---preserve the Gaussian character of these states, thereby simplifying both theoretical analysis and practical implementation. This framework underlies many CV-QKD protocols and quantum teleportation schemes, providing robust security proofs \cite{28,29,30,31} and enabling precise error correction \cite{30,32}. However, while Gaussian protocols benefit from ease of manipulation and scalability, they face inherent limitations such as the impossibility of distilling entanglement using only Gaussian operations. Consequently, integrating non-Gaussian \cite{33,34} elements remain an active area of research to enhance the capabilities of quantum networks. The synergy between Gaussian quantum information theory and photonic integrated circuit technology continues to drive advancements in secure, high-speed quantum communication.

Recent advances in machine learning have also begun to influence the field of CV-QKD \cite{35,36}. Data-driven algorithms, including deep neural networks and reinforcement learning models, are being employed to optimize system performance and enhance security. By analyzing extensive datasets of channel statistics and noise fluctuations, these machine learning techniques enable real-time adaptive control of modulation parameters~\cite{37}, error correction protocols~\cite{38}, and channel estimation. This results in improved secret key rates and extended transmission distances. Furthermore, machine learning-assisted CV-QKD systems can effectively detect anomalous patterns indicative of eavesdropping attempts \cite{39,40}, thus adding an extra layer of security. The integration of these advanced data analytics not only streamlines the operational aspects of quantum communication but also accelerates the transition from experimental setups to robust, large-scale deployments.

Beyond quantum communication, squeezed light finds extensive applications in quantum computation and quantum information processing. Squeezed qubits, formed using squeezed light~\cite{41}, are quantum states in which one observable, typically amplitude or phase, exhibits reduced noise compared to the surrounding quantum state. By reducing noise in one quadrature while maintaining quantum uncertainty, squeezed qubits enable enhanced measurement precision and improved fidelity in quantum operations. Squeezed light qubits offer several advantages in quantum computation. They enhance the accuracy of quantum gates and measurements, increasing the reliability of quantum operations. Additionally, they improve resilience against certain types of noise and decoherence, facilitating prolonged quantum information storage and processing. Moreover, squeezed light qubits enable novel quantum algorithms and protocols that exploit their unique properties.

The history of digital computation demonstrates that major advancements arise from increasing transistor density in integrated circuits, leading to higher processing speeds and energy efficiency. Similarly, photonic integrated circuits (PICs) hold promises for revolutionizing photonic technologies. While bulk photonics has been well-established, quantum photonic integrated circuits (qPICs) are gaining traction in quantum technologies. The use of squeezed light in qPICs presents exciting opportunities in quantum computation and simulation, quantum sensing and metrology, and quantum communication.

Quantum computation and simulation have significantly benefited from the integrated generation of squeezed light states, enhancing Gaussian boson sampling and quantum walk experiments \cite{42,43}. These advances mitigate the probabilistic nature of parametric downconversion sources~\cite{44}. The demonstration of quantum supremacy using squeezed light sources \cite{45,46,47} underscores its potential for quantum computing applications. Additionally, squeezed states help counteract certain types of quantum noise in qPICs \cite{48,49,50}. Quantum sensing and metrology have experienced rapid growth, with squeezed light playing a vital role in improving measurement precision. The key advantage of squeezed light over coherent light~\cite{51} lies in the enhanced accuracy of one quadrature while sacrificing precision in another. This characteristic enables highly sensitive measurements, as evidenced by the use of squeezed light in improving LIGO's gravitational wave detection capabilities \cite{52,53}. Quantum communication is integral to QKD and quantum networking, offering robust security and compatibility with integrated photonic platforms. The continued integration of squeezed states in quantum communication systems is expected to further advance secure quantum networks.

Recent research has also focused on the development of hybrid QKD systems that combine both DV-QKD and CV-QKD protocols~\cite{54}. This integrated approach seeks to harness the strengths of both methods: the long-distance transmission capabilities of DV-QKD and the high key rates and infrastructural compatibility of CV-QKD. Efforts to refine error correction, improve signal processing techniques, and develop advanced quantum noise reduction methods are accelerating progress toward resilient, large-scale quantum networks. Furthermore, innovative designs in qPICs, which embed sources of squeezed light, modulators, and detectors on a single chip, are poised to revolutionize the practical deployment of quantum communication systems. These integrated platforms promise not only enhanced performance and stability but also greater scalability and cost-effectiveness, paving the way for a future quantum internet that seamlessly connects quantum processors and sensors over global distances.

Section~\ref{sec2} of this paper introduces the concept of squeezed states of light, discussing their generation and fundamental properties. Section~\ref{sec3} explores continuous-variable quantum teleportation, covering the principles, experimental approaches, and integration of essential components. In Section~\ref{sec4}, we provide a comprehensive survey of continuous-variable quantum key distribution (CV-QKD), including its fundamental principles, experimental demonstrations, and recent progress in chip-scale integration. Section~\ref{sec5} discusses feasibility and challenges in CV-QKD, addressing issues such as squeezed light generation, photonic integration, photon loss, and security concerns. Finally, Section~\ref{sec6} concludes the paper by summarizing key findings and outlining future research directions.

\section{Squeezed State of Light}\label{sec2}

A squeezed state \cite{55,56} of light is a unique quantum light state that can be understood through its connection to the quantum harmonic oscillator. Here, the oscillation of the electromagnetic field is governed by two non-commuting quadrature operators, analogous to position and momentum in a mechanical system. For coherent light, these quadratures exhibit equal uncertainty, leading to a circular uncertainty region in phase space with a minimum product of uncertainties, 1/2, as dictated by the Heisenberg uncertainty principle. When the wave functions $\psi (x)$ and $\varphi (p)$ are Gaussian---given by $\psi (x)\propto e^{{-(x-\alpha )}^2}$ and $\varphi (p)\propto e^{{-(p-\beta )}^2}$---this minimum uncertainty product for position (\textit{x}) and momentum (\textit{p}) reaches 1/4. Here, \textit{$\alpha $} and \textit{$\beta $} denote the average values of \textit{x} and \textit{p}, respectively.

In coherent light, the photon number statistics follow a Poisson distribution, and the state is inherently non-deterministic. In applications requiring precise time-domain measurements, such as quantum integrated photonics, this uncertainty becomes a significant limitation, particularly when using vacuum states or single-photon states. Such scenarios demand a more controlled quantum light source---a squeezed state---to mitigate measurement uncertainty.

\begin{figure}[H]%F1
\centering
\includegraphics[scale=0.8]{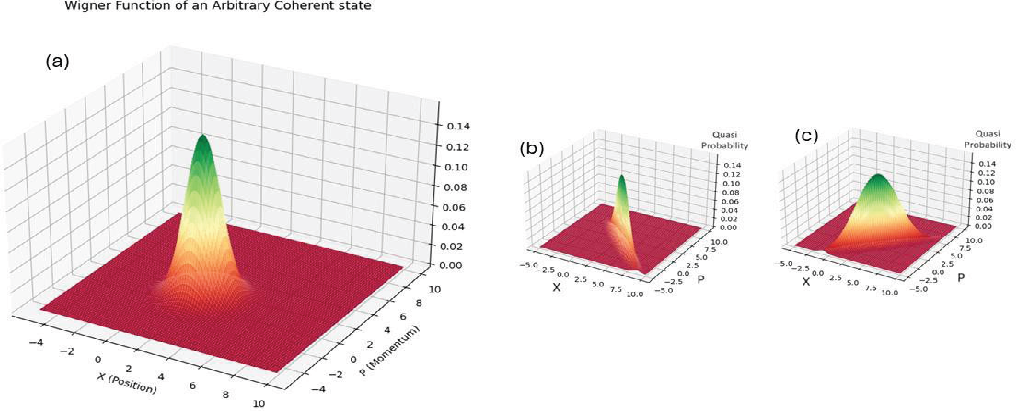}
\caption{(\textbf{a}) An arbitrary coherent state (\textbf{b}) is the amplitude-squeezed state of light and (\textbf{c}) is the phase-squeezed state of light. The quasi-probability of the perpendicular axis displays the Wigner function value.}
\label{fig1}	
\end{figure}

Squeezed light reduces the uncertainty in one quadrature (e.g., amplitude or phase) to below 1/2 by deforming the circular uncertainty region into an ellipse. This reduction enhances precision in specific quadrature measurements while increasing uncertainty in the orthogonal quadrature, consistent with the uncertainty principle. Figure~\ref{fig1}b and c illustrate the amplitude-squeezed and phase-squeezed scenarios, respectively. By heterodyning a squeezed field with coherent light, it is possible to achieve reduced noise suitable for precision measurements in quantum technologies. Furthermore, squeezed light exhibits sub-Poissonian photon statistics, making it advantageous for low-noise quantum systems.

Squeezed states are generated using nonlinear optical processes such as parametric down-conversion, parametric oscillation, and twin-beam generation. These mechanisms form the basis of practical squeezed light sources for applications such as Mach-Zehnder interferometers (MZIs) and photonic arrays~\cite{57}.

After the quantization of electromagnetic waves (photons), the relevant quantities are the position quadrature (\textit{x}) and the momentum quadrature (\textit{p}) of the photon. Squeezed states are characterized by a reduction in the uncertainty of either the electric field's magnitude or its phase, as illustrated in Figure~\ref{fig1}b and Figure~\ref{fig1}c, respectively. The squeezed state operator is defined as:
\begin{equation}\label{eq1}
\hat{S}\left(\xi \right)={{exp}}\ exp\ \left[\frac{\left(\xi {\hat{a}}^2-{\xi }^*{\hat{a}}^{\dagger 2}\right)}{2}\right]
\end{equation}

In this context, $\hat{S}(\xi )$ represents a one-mode squeezed light operator, and $\hat{a}\ $and ${\hat{a}}^{\dagger }$ are the annihilation and creation operators, respectively, with $\xi $ being the squeezing parameter where $\xi =re^{i\vartheta}$. Meanwhile, the two-mode squeezed light operator is:
\begin{equation}\label{eq2}
{\hat{S}}_2\left(\xi \right)={{exp}}\ exp\left[\left(-\xi \hat{a}\hat{b}+{\xi }^*{\hat{a}}^{\dagger }{\hat{b}}^{\dagger }\right)\right]
\end{equation}

In the case of Eq.~\eqref{eq2}, all of the parameters are identical to those in the previous equation, with the exception of the $\hat{b}$ (and ${\hat{b}}^{\dagger }$) operator, which serves as the annihilation (and creation) operator for the second mode. You can refer to~\cite{56} for a complete derivation and discussion on squeezing operators.

\begin{figure}[t]%F2
\centering
\includegraphics[scale=0.8]{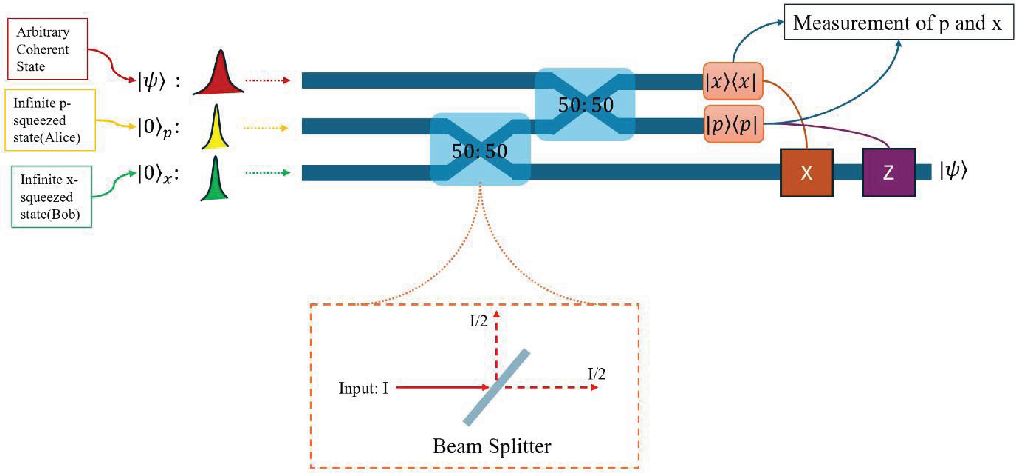}
\caption{Illustration of continuous-variable quantum teleportation. An arbitrary coherent state $|\psi \rangle$ (top rail) is to be teleported from Alice to Bob. Two infinitely squeezed vacuum states (middle and bottom rails) serve as the shared resource between them. First, these two squeezed states pass through a 50:50 beam splitter (left 50:50 block) to create an entangled EPR-like pair, with one mode held by Alice and the other mode sent to Bob. Next, Alice couples her unknown state $|\psi \rangle$ with her half of the entangled pair using another 50:50 beam splitter (right 50:50 block). She then performs homodyne measurements of both the amplitude (\textit{x}) and phase (\textit{p}) quadratures on the combined output. The measured values, $\langle \textit{x}\rangle$ and $\langle\textit{p}\rangle$, are sent via a classical channel to Bob, who applies corresponding displacement operations (the X and Z boxes) to his mode. This ``corrects'' his half of the entangled pair into an identical copy of $|\psi \rangle$. The inset (lower box) shows the 50:50 beam splitter's action, splitting or combining two input modes in equal proportions.}
\label{fig2}	
\end{figure}

\section{Continuous Variable Quantum Teleportation}\label{sec3}

\subsection{Quantum Teleportation}

Quantum teleportation~\cite{58} enables the transfer of an undisclosed quantum state between distant qubits or qumodes. It relies on classical communication and quantum entanglement, playing a crucial role in quantum information protocols. It also has applications in distributed information processing for quantum computation.

Quantum teleportation circuits operate on a consistent fundamental principle. Two observers, Alice and Bob, situated apart, share a maximally entangled quantum state~\cite{59}. This state can be one of the four Bell states for discrete variables or a maximally entangled state with a fixed energy level for continuous variables. They also have access to a classical communication channel as shown in Figure~\ref{fig2}.

To transfer an unknown state from Alice to Bob, Alice conducts a joint measurement on her portion of the entangled state and the unknown state. This measurement involves projecting onto the Bell basis. Alice subsequently communicates the measurement results to Bob. Armed with this information, Bob can convert his half of the entangled state into an accurate copy of the original unknown state. This transformation may entail a conditional phase flip for qubits or displacement for qumodes.

The initial concept of quantum teleportation focused on discrete-variable quantum communication with qubits. However, its versatility also applies to continuous-variable qumodes, especially when they are spatially distant. In Figure~\ref{fig2}, we can see the circuit specifically designed for continuous-variable quantum teleportation, where BS stands for beam splitter, and X and Z are position displacement and momentum displacement which are defined as $\hat{D}\left(x_0\right)=e^{-\frac{i}{\hbar }x_0\hat{p}}$ and $\hat{D}\left(p_0\right)=e^{\frac{i}{\hbar }p_0\hat{x}}$, respectively, to recover the state $|\psi \rangle$ exactly (The impact of these operators are $\hat{D}\left(x_0\right)\psi (x)=\psi (x-x_0)$ and $\hat{D}\left(p_0\right)\varphi (x)=\varphi (p-p_0)$, respectively).

The Strawberry Fields Library in Python was used to effectively simulate the squeezed states and their quantum state evolution through the proposed algorithm. This library facilitates the simulation of such photonic devices for quantum applications.

Assuming we aim to teleport an arbitrary coherent state such as Figure~\ref{fig1}a. As shown in Figure~\ref{fig2}, The entangled states of Bob and Alice should be a combination of states as depicted. They are considered entangled due to the beam splitter transformation applied to the initial infinite squeezed states. By performing the third and fourth steps on the states, Bob will achieve the same state as depicted in~\cite{60}.

\begin{figure}[H]%F3
\centering
\includegraphics[scale=0.85]{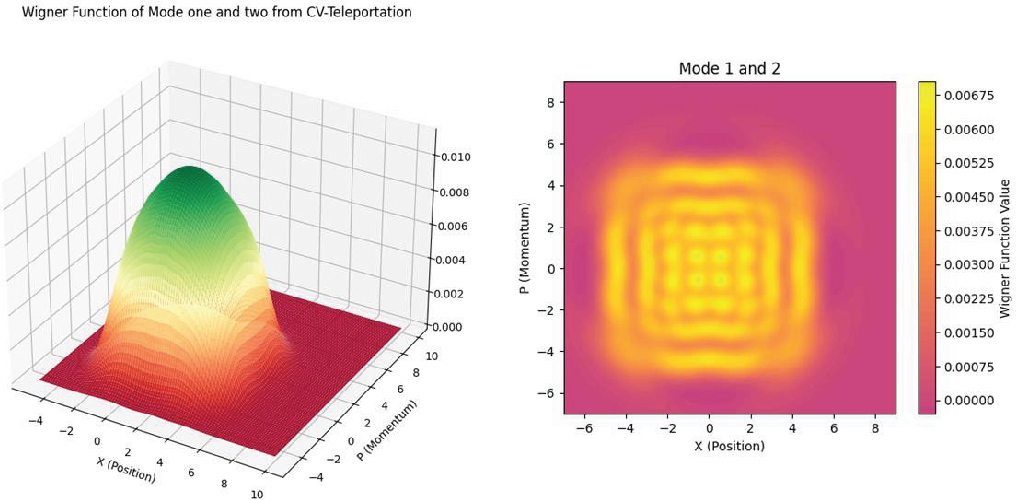}
\caption{Wigner function of quadrature diagram of Bob and Alice entangled states after going through beamsplitters.}
\label{fig3}	
\end{figure}

The process of continuous variables quantum teleportation depicted in Figure~\ref{fig2} (for simplicity we call the coherent state ate the beginning of the circuit mode 0 and Alice and Bob become mode 1 and mode 2 respectively) can be paraphrased as follows:
\begin{enumerate}
\item[(a)] Initially, the qumodes ${|0\rangle}_p$ and ${|0\rangle}_x$ (Alice and Bob) are set up as infinitely squeezed vacuum states in momentum and position space, respectively. First, we express the meaning of vacuum squeezed state which is:
\begin{equation}\label{eq3}
|\xi \rangle=\hat{S}\left(\xi \right)|0\rangle
\end{equation}
where $\hat{S}(\xi)$ is defined as Equation~\eqref{eq1}. So, one can calculate the squeezed position wave function as follows~\cite{61}:
\begin{equation}\label{eq4}
\langle\xi \rangle={\psi }_r\left(x\right)={\left(\frac{\lambda }{\pi }\right)}^{\frac{1}{4}}{{exp}}exp\ \left(-\frac{\lambda x^2}{2}\right)
\end{equation}
where $\lambda =e^{2r}$, which is called the squeezing parameter. By taking the squeezing parameter into infinite the squeezed position wave function becomes Dirac delta function and the uncertainty in position vanishes. Moreover, same calculations could be done for infinite momentum vacuum squeezed state which is defined as $\langle\xi \rangle={\psi }_r\left(p\right)={(\frac{1}{\lambda \pi })}^{1/4}exp(-\frac{x^2}{2\lambda })$.
\item[(b)]  Subsequently, mode 1 and mode 2 (Alice and Bob) become maximally entangled using a 50-50 beamsplitter, represented as:
\begin{equation}\label{eq5}
BS\left(\frac{\pi }{4,0}\right)\left(|0\rangle_p\bigotimes |0\rangle_x\right)
\end{equation}
which $BS(\theta ,\varphi )$ is defined as $\left(cos\theta\ -e^{-i\varphi }sin\theta \ e^{-i\varphi }sin\theta\ cos\theta\ \right)$ where after this, the qumodes are spatially separated, with Alice holding the state as shown in Figure~\ref{fig3}. They are linked through classical communication channels $c_0$ and $c_1$.

\item[(c)] To teleport her unknown state $|\psi \rangle$ to Bob, Alice must carry out a projective measurement of her entire system using the maximally entangled basis states. This is accomplished by entangling $|\psi \rangle$ and $|q_1\rangle$ via another 50-50 beamsplitter which its Wigner function is shown in

\item[(d)] The results of the measurements are sent to Bob, who uses them to execute a position displacement (conditional on the \textit{x} measurement) and a momentum displacement (conditional on the p measurement). These displacements enable Bob to accurately retrieve the transmitted state $|\psi \rangle$.
\end{enumerate}

\subsection{Integration of Elements}

\begin{figure}[H]%F4
\centering
\includegraphics[scale=0.85]{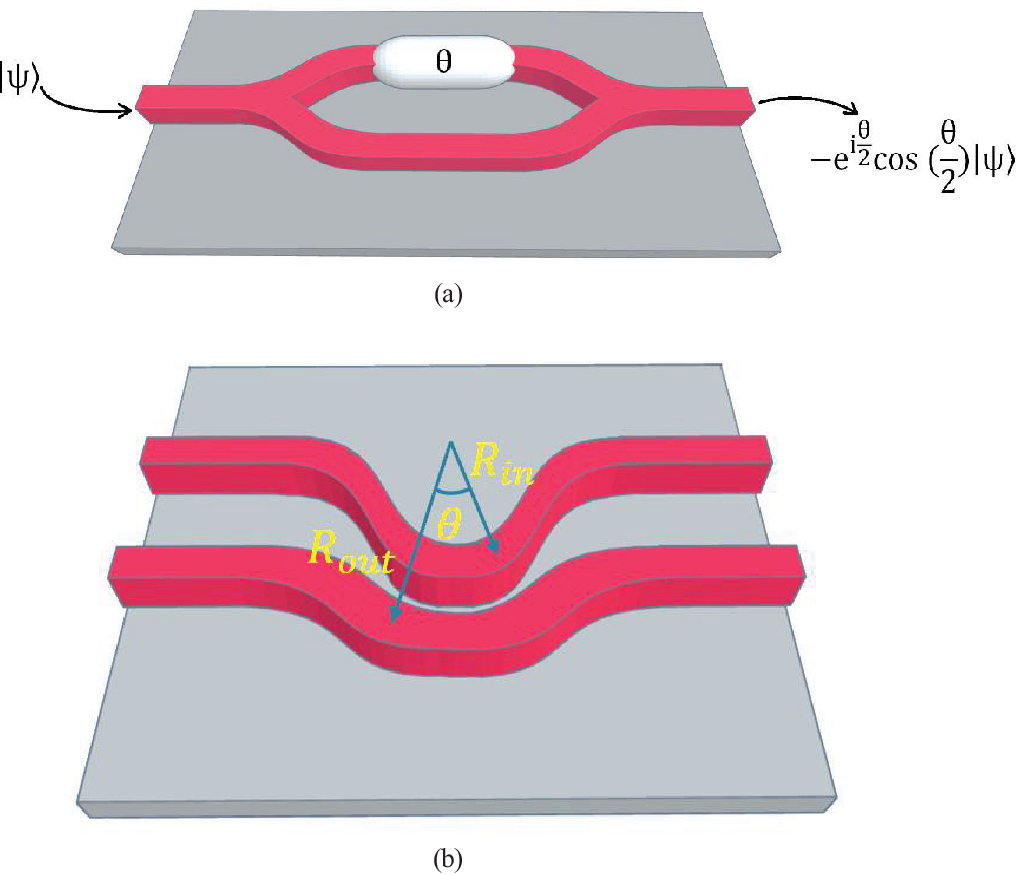}
\caption{(\textbf{a}) A modified integrated unit cell inspired by the Mach-Zehnder Cell. It can act as an arbitrary Measurement gate for identifying state's position or momentum by changing the variable $\theta $ of the phase shifter. (\textbf{b}) Bent directional coupler with labeled parameters: $R_{in}$ and $R_{out}$ represent the inner and outer bending radii, respectively, while $\theta $ indicates the bending angle. The structure is designed to facilitate efficient light coupling between waveguides while minimizing losses and cross-talk effects.}
\label{fig4}	
\end{figure}

As pointed out in~\cite{62}, any unitary transformation on individual qubits can be depicted as a series of three rotations, one around the $\hat{y}$ axis and two around the $\hat{z}$ axis. Motivated by this fact, a Mach-Zehnder integrated cell is appropriate for implementing an arbitrary single-qubit unitary transformation~\cite{63}. An alteration to the Mach-Zehnder interferometer cell can be made to align with the structure illustrated in Figure~\ref{fig2}. The updated integrated cell is shown in Figure~\ref{fig4}a.

\newpage

\begin{figure}[t]%F5
\centering
\includegraphics[scale=0.85]{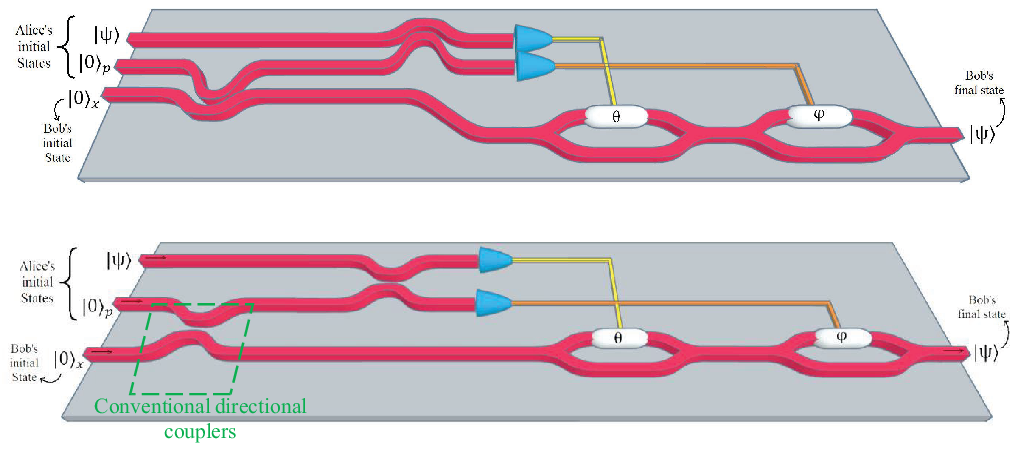}
\caption{A 3D representation of a Photonic Integrated Circuit (PIC) designed to implement quantum teleportation in a continuous quantum variable circuit. The input states are introduced into the circuit from the left where (\textbf{a}) is with bent directional couplers and (\textbf{b}) is with conventional directional couplers, and at the conclusion (right side), Bob will possess Alice's initial state of $\left|\psi \right\rangle $. The two blue cones symbolize homodyne detectors that measure the average value of arbitrary quadrature components. The outcomes of these measurements are utilized as controls for X and Z which are position displacement and momentum displacement, respectively, to recover the state $\left|\psi \right\rangle $ exactly.}
\label{fig5}	
\end{figure}

Elements of the structure shown in Figure~\ref{fig2} can be easily reconfigured using two of these newly adapted cells, two homodyne detectors, and two directional couplers. The resulting circuit will resemble Figure~\ref{fig5}. The two directional couplers are necessary to create entanglements between the states of Alice and Bob. They function similarly to the beamsplitters in Figure~\ref{fig2}. This type of circuit will also require two homodyne detectors to measure the average value of arbitrary quadrature components. It's important to note that in this setup, the quantum states are not formed using single photons, but rather continuous variable quantum states that have the capability to be squeezed. Moreover, bent directional couplers \cite{64,65} as depicted in Figure~\ref{fig4}b play a pivotal role in the design and optimization of integrated photonic circuits, particularly within Mach-Zehnder interferometers (MZIs). These couplers facilitate efficient light propagation and interaction by guiding optical signals through curved paths with minimal losses.

Their inclusion in the system enhances the compactness and scalability of the overall architecture, which is essential for quantum communication systems. Furthermore, the use of bent directional couplers ensures precise splitting and combining of optical signals, thereby maintaining the coherence and stability critical for continuous variable quantum key distribution (CV-QKD). By enabling tighter integration of components, these couplers support the development of high-performance and miniaturized photonic devices for secure data transmission which can be seen as a new avenue.

\section{Continuous Variable Quantum Key Distribution}\label{sec4}

\subsection{Basic Principles of CV-QKD}

Quantum Key Distribution (QKD) has undergone significant advancements with the development of Continuous Variable (CV) protocols, starting from Ralph's pioneering proposal to encode key information using the amplitude and phase quadratures of the light field. Initially, discrete modulation schemes focused on binary-encoded keys \hbox{\cite{66,67,68}}. However, continuous modulation protocols, such as Gaussian modulation, soon emerged, utilizing coherent and squeezed states of light \cite{69,70,71}.

Modulation plays a central role in CV-QKD~\cite{72}, encoding classical information into quantum states for secure key exchange. In Gaussian modulation, Alice prepares coherent or squeezed states by modulating their quadrature components---amplitude and phase---using Gaussian-distributed random variables. Electro-optic modulators driven by high-speed random number generators achieve these modulations, enabling high-dimensional encoding that improves key rates and resilience against noise.

For discrete modulation, Alice selects from a finite set of states, such as quadrature phase-shift keying (QPSK) or other constellation formats. While this approach simplifies state preparation and error correction, it often requires more sophisticated security proofs.

Alice transmits the modulated quantum states over an insecure quantum channel---optical fiber or free space---to Bob, who measures the quadratures using homodyne or heterodyne detection. Homodyne detection involves Bob actively selecting and measuring either the amplitude or phase quadrature, leading to data sifting where half of the data is discarded to ensure matched bases. Heterodyne detection, by contrast, measures both quadratures simultaneously, eliminating sifting and enhancing the key rate.

The no-switching protocol utilizes heterodyne detection with Gaussian-modulated coherent states, simplifying implementation and improving efficiency. Furthermore, protocols leveraging Einstein--Podolsky--Rosen (EPR) entangled states demonstrate that modulation can be avoided altogether, relying on quantum correlations instead~\cite{73}.

Modulation techniques have evolved to address challenges such as excess noise, channel loss, and implementation complexity~\cite{74}. Doubly modulated coherent states reduce inefficiencies by requiring only Alice to discard half of her data during sifting. Additionally, modulating entangled states of light has improved robustness to channel noise, resulting in higher key rates and extended distribution distances \cite{75,76}.

Digital signal processing (DSP) and hardware advancements have further enhanced modulation precision and system performance. Modern CV-QKD systems utilize high-speed modulators and advanced reconciliation algorithms to maximize key rates and minimize errors. For example, heterodyne-based Gaussian-modulated coherent state protocols eliminate active basis choice and sifting, enhancing the key rate \cite{77,78}.

The secret key rate in CV-QKD is given by:
\begin{equation}\label{eq6}
K=\beta I\left(A:B\right)-\chi \left(E\right)
\end{equation}
where \textit{$\beta $} is Reconciliation efficiency which is in [0,1], $I(A:B)$ is Mutual information between Alice and Bob, computed~as:
\begin{equation}\label{eq7}
I\left(A:B\right)=H\left(A\right)+H\left(B\right)-H\left(A,B\right)
\end{equation}
where $H(\cdot)$ represents the Shannon entropy, and $\chi (E)$ is Holevo bound, representing Eve's maximum accessible information.

For Gaussian-modulated coherent states, the mutual information is:
\begin{equation}\label{eq8}
I\left(A:B\right)=\frac{1}{2}\left(1+\frac{V_s}{V_N}\right)
\end{equation}
where $V_s$ is the signal variance, and $V_N$ is the noise variance.

The Holevo bound is calculated using the symplectic eigenvalues of the covariance matrices describing the quantum states. It represents the upper bound of information accessible to an eavesdropper.

By leveraging quantum correlations, CV-QKD protocols utilizing EPR entangled states have demonstrated improved robustness and higher key rates, even without direct signal modulation~\cite{73}. For instance, a protocol using squeezed states and heterodyne detection proposed in 2009 achieved higher secret key rates than earlier Gaussian protocols~\cite{79}.

CV-QKD can be implemented in two frameworks: the prepare-and-measure (PM) scheme and the entanglement-based (EB) scheme. The PM scheme is typically simpler to implement, while the EB scheme offers a unified theoretical framework for key rate calculations. These two approaches are equivalent for Gaussian protocols \cite{80,81}.

In summary, advancements in modulation methods, hardware, and theoretical models have enabled CV-QKD to achieve higher key rates, extended distribution distances, and enhanced robustness against noise, solidifying its position as a viable solution for secure quantum communication.

\subsection{Experimental Implementation of CV-QKD}

\begin{figure}[t]%F6
\centering
\includegraphics[scale=0.85]{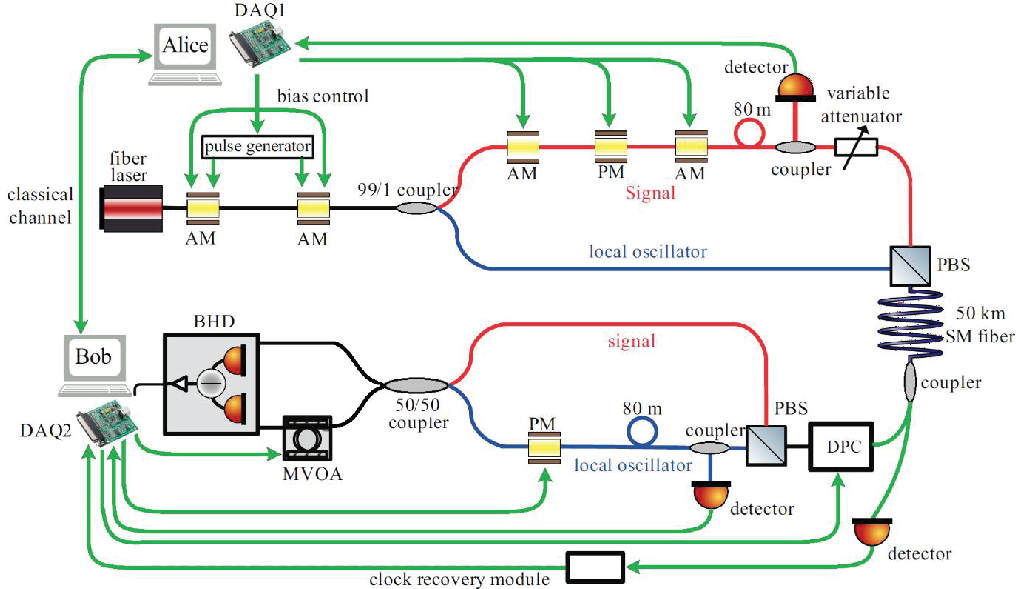}
\caption{This is the experimental configuration proposal for a 50-km Continuous Variable Quantum Key Distribution (CV QKD). In this setup, DAQ refers to the data acquisition module, AM is the amplitude modulator, PM stands for the phase modulator, and PBS is the polarizing beam splitter. DPC is the dynamic polarization controller, MVOA denotes the motor variable optical attenuator, and BHD is the balanced homodyne detector.}
\label{fig6}	
\end{figure}

Figure~\ref{fig6} illustrates an experimental setup for long-distance Continuous Variable Quantum Key Distribution (CV QKD)~\cite{22}. This process is achieved by utilizing two cascaded intensity modulators to generate high-extinction-ratio optical pulses with a width of 100~ns and a repetition rate of 500 kHz from a 1550-nm continuous-wave (CW) single-frequency fiber laser. These optical pulses are then split into a weak signal field and an intense local oscillator (LO) using a fiber coupler (99/1). The pulsed coherent states undergo bivariate Gaussian modulation in the phase space through a combination of an intensity modulator and a phase modulator. The signal and LO are directed through a 50-km single-mode fiber spool using time multiplexing and polarization multiplexing, achieved by an \hbox{80-m} fiber and a polarizing beam splitter (PBS), respectively. When Bob receives the quantum states sent by Alice, he randomly measures either the amplitude or phase quadrature of the signal using a pulsed balanced homodyne detector (BHD). The random basis switch and relative phase locking between the signal and LO are implemented with a phase modulator in the LO path. To accurately measure the BHD output's peak voltage and synchronize the system, a small portion of the LO beam is extracted to recover the system clock, which is then precisely delayed to the desired value.

In another study~\cite{82}, Roberts et al. address a significant challenge in high-rate decoy-state Quantum Key Distribution (QKD) systems: the instability of intensity modulators (IMs), which can lead to security vulnerabilities known as side channels. These side channels arise when the intensity of a light pulse depends on the preceding pulse, a phenomenon termed the ``patterning effect.'' This effect can inadvertently leak information, compromising the security of the QKD system. To mitigate this issue, the researchers propose the use of a tunable extinction ratio Sagnac-based intensity modulator. This design offers several advantages:
\begin{enumerate}
\item[A.] Patterning Effect Mitigation: By operating at specific modulation points, the Sagnac-based IM ensures that the output intensity of each pulse is independent of previous pulses, effectively eliminating the patterning effect.

\item[B.] Enhanced Stability: The inherent design of the Sagnac interferometer provides immunity to direct current (DC) drifts, ensuring consistent performance without the need for frequent recalibration.

\item[C.] Tunable Extinction Ratios: The modulator allows for the selection of low extinction ratios, enabling the faithful reproduction of random decoy state patterns essential for secure QKD operations.
\end{enumerate}

By integrating this Sagnac-based IM into QKD systems, a practical solution to enhance security and performance is achieved, addressing critical issues associated with traditional intensity modulators.

Excess noise significantly impacts the performance of the system, affecting both the secure key rate and transmission distance. Ideally, excess noise results solely from Eve's eavesdropping behavior. However, in practice, technical noises may arise even without eavesdropping. These technical noises include LO pulse leakage due to finite optical pulse extinction ratios, quantum state preparation, non-ideal phase locking between the LO and signal, nonlinear optical effects in the fiber, and BHD stability.

In their comprehensive review, Zhang et~al.~\cite{22} emphasize the pivotal role of laser stability in Continuous-Variable Quantum Key Distribution (CV-QKD) systems. They highlight those fluctuations in the laser's frequency and intensity can introduce excess noise, which adversely affects the system's performance and security. To mitigate these issues, the authors recommend implementing precise temperature control and effective vibration isolation. By maintaining a stable operating environment, these measures can significantly reduce frequency drift and intensity noise, thereby enhancing the reliability and security of CV-QKD systems.

While Eve typically cannot manipulate these noises effectively, they can vary over time, necessitating real-time monitoring to characterize system parameters. In practical QKD systems, such noises are often attributed to Eve's attack when assessing security.

Although some possible challenges were mentioned in this part, further and deeper discussions about security concerns of CV-QKD and possible solutions such as measurement-device-independent~\cite{83} methods are covered in a further section of this paper.

\subsection{CV-QKD on an Integrated Chip}

As discussed in the Introduction, fiber-based DV-QKD has been successfully demonstrated over ultra-low-loss fibers, achieving key rates at the kbps level for distances up to 100 km. Several DV-QKD protocols, including encoding based on spatial dimensions~\cite{84}, photon polarization~\cite{85}, and time bins~\cite{86}, have been realized on silicon wafers. For on-chip photon detection, superconducting nanowire-based single-photon detectors with detection efficiencies of up to 90\% are integrated \cite{87,88}. Notably, CV-QKD is particularly suitable for photonic chip integration due to its compatibility with existing telecom technologies \cite{24,89}.

In 2013, a fiber-based Continuous Variable Quantum Key Distribution (CV-QKD) system achieved a secure key rate of approximately 1 kbps over an 80 km transmission distance~\cite{90}. By effectively managing excess noise, the distance was subsequently extended to over 100 km~\cite{91}.

Recently, there have been reports of on-chip quantum entropy sources based on detecting vacuum fluctuations~\cite{92} and phase fluctuations \cite{93,94,95}. Notably, a chip-based homodyne detector demonstrated a gain of 4.5 kVA${}^{-1}$ with a 150-MHz bandwidth~\cite{92}.

Despite these advancements, integrating all necessary components for CV-QKD systems into a single chip presents significant challenges due to the diverse material requirements and fabrication processes for each component~\cite{96}. One way to overcome this challenge is by modular integration. This strategy involves fabricating distinct functional blocks---such as modulators, detectors, and couplers---separately, and subsequently interconnecting them. This method simplifies fabrication, enhances scalability, and facilitates easier upgrades and maintenance~\cite{97}.

Existing techniques supporting modular integration include heterogeneous and hybrid integration. Heterogeneous integration combines different materials, such as III-V semiconductors with silicon photonics~\cite{98}, enabling the integration of active components like lasers and amplifiers onto silicon-based platforms. Hybrid integration involves the optical or electrical coupling of separate chips, each optimized for specific functions, to form a cohesive photonic system. Packaging techniques, such as photonic wire bonding~\cite{99}, facilitate seamless communication between these integrated modules by providing low-loss interconnections.

Applying modular integration to CV-QKD architectures can significantly enhance system performance. For instance, components like Mach-Zehnder interferometers and bent directional couplers can be fabricated independently and then interconnected, allowing for optimized performance of each module. This approach not only reduces fabrication complexity but also allows for the selection of the most suitable materials and fabrication processes for each component, thereby improving the overall efficiency and reliability of the CV-QKD system~\cite{97}.

\section{Feasibility and Challenges}\label{sec5}

While the utilization of squeezed light offers unparalleled precision, it's not without its compromises. Implementing a squeezed state of light requires careful thought, from the challenges associated with producing these states~\cite{100}, to the design and manufacture of compatible Photonic Integrated Circuits (PICs). In this section, we will examine three of these issues and the measures taken to address them.

\subsection{Methods to Generate Squeezed Light}

\subsubsection{Reducing Pump Fluctuation in Semiconductor Laser}

This holistic approach considers both the thermal fluctuations in the flow of minority carriers and the quantum-mechanical generation-recombination noise they generate~\cite{101}. To gain a deeper understanding of this phenomenon, it's important to examine the injection-current-pumped semiconductor laser. In this type of laser, the carrier-injection process is governed by a similar effect. The rate of carrier injection, referred to as the junction current, is dictated by the forward-biased voltage applied across a p-n junction. This voltage counteracts the inherent potential within a depletion layer, thus allowing the carrier-diffusion process to take precedence over the reverse-directed drift process.

It's noteworthy that this modulation effect on the junction voltage, induced by a series resistance, plays a pivotal role in controlling the process of minority carrier injection into the active layer. This view of the injection process and its control is crucial for understanding the behavior of semiconductor lasers.

\subsubsection{Nonlinear Optics}

\begin{figure}[t]%F7
\centering
\includegraphics[scale=0.85]{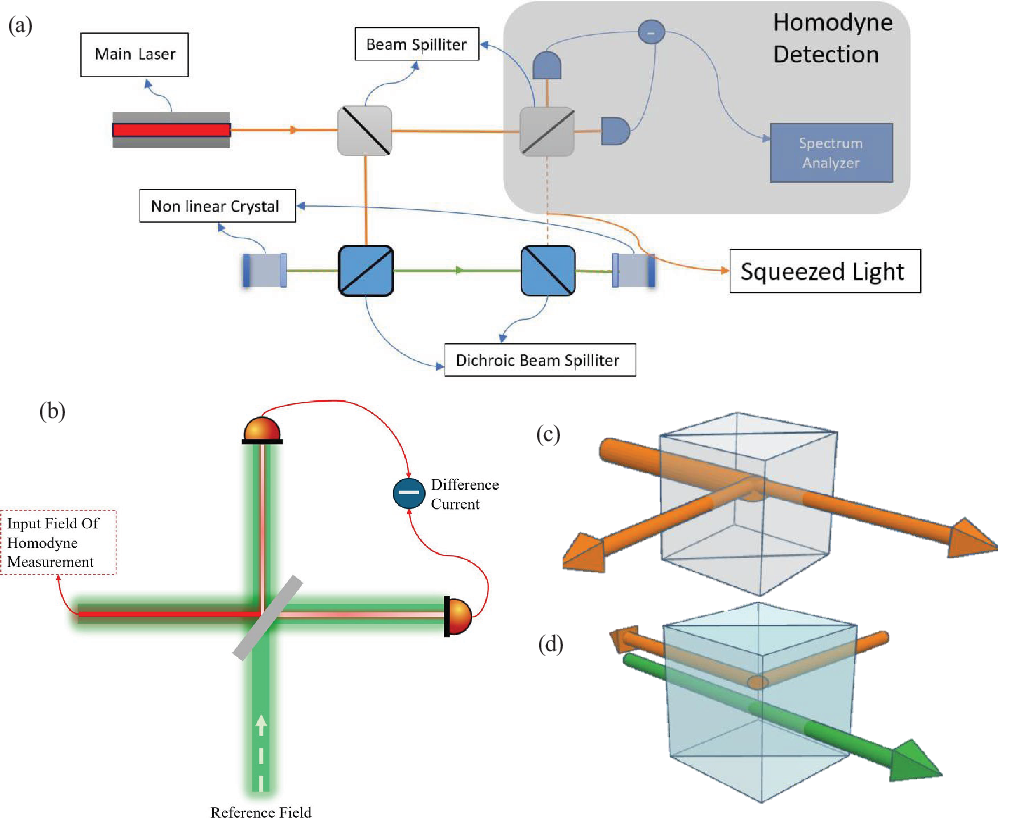}
\caption{(\textbf{a}) A detailed optical setup for generating and measuring squeezed light. An infrared laser (main laser) undergoes second-harmonic generation (SHG) via a nonlinear crystal, converting it into a green pump beam. This pump beam is then used to drive an optical parametric amplifier (OPA), which generates squeezed infrared light. Dichroic and beam splitters direct different frequency components through the system. The output squeezed state is analyzed using homodyne detection, where the interference pattern is recorded by a spectrum analyzer. (\textbf{b}) A conceptual diagram of homodyne detection. The reference field (local oscillator) interferes with the squeezed field at a beam splitter, producing two outputs detected by photodiodes. The difference in photocurrent reveals quadrature noise properties of the squeezed state. (\textbf{c}) and (\textbf{d}) illustrate interactions at beam splitters. (\textbf{c}) represents a standard beam splitter operation where input beams mix and split into new paths. (\textbf{d}) shows dichroic beam splitting, where different wavelength components are separated, enabling selective manipulation of infrared and green light in the experiment.}
\label{fig7}	
\end{figure}

Nonlinear optics is an area that investigates the interaction between light and materials exhibiting nonlinear characteristics. One fascinating application of nonlinear optics is the creation of squeezed states of light. While one method of squeezing light involves suppressing pump fluctuation in semiconductor lasers, a more prevalent approach typically involves directing a laser beam through a nonlinear crystal or medium. The nonlinear characteristics of the material trigger a nonlinear interaction in the light, leading to the production of squeezed states. This technique has demonstrated more promising outcomes in experiments compared to the previous method (Figure~\ref{fig7}).

Various strategies are employed in nonlinear optics to produce squeezed states, such as parametric down-conversion \cite{102} and four-wave mixing \cite{103,104}. These methods involve altering the properties of the input laser beam, like its intensity or frequency, to provoke the desired nonlinear interaction and create the squeezed state.

\subsubsection{Integrated Micro-Ring Resonators}

In addition to their numerous applications, ring resonators can be made from various materials that display nonlinear reactions to high-intensity light. This nonlinearity facilitates frequency modulation processes such as four-wave mixing and spontaneous parametric down-conversion, which produce photon pairs~\cite{105}. These processes can be further employed to squeeze light beams in photonic integrated circuits \cite{103,104}, offering significant benefits for a wide range of areas, from telecommunications to both classical and quantum computing.

\subsection{Appropriate Integration for Squeezed Light}

The optimization of Photonic Integrated Circuits (PICs) for the application of squeezed light presents a challenge, as most PICs are specifically designed for coherent light or single photons. A significant hurdle is integrating the squeezed light source with other components of the PIC. Despite these challenges, efforts are being made to address them, such as an experimental chip designed by Xanadu~\cite{106}. This device is a programmable nanophotonic chip capable of performing quantum circuit tasks using a squeezed light beam.

\begin{figure}[H]%F8
\centering
\includegraphics[scale=0.85]{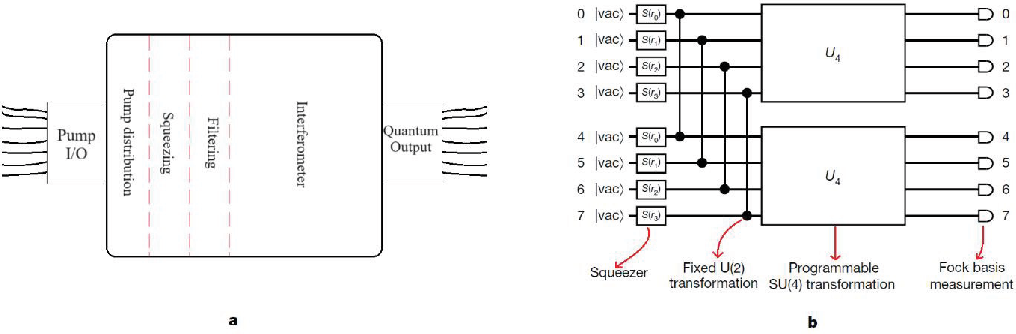}
\caption{(\textbf{a}) The chip is being reintroduced, as shown in a micro-graph of the actual device. (\textbf{b}) The provided quantum circuit diagram serves as a comparable representation of the photonic hardware, emphasizing its functional features similar to~\cite{69}.}
\label{fig8}	
\end{figure}

As per Figure~\ref{fig8}a, the core component of Xanadu's programmable device is a photonic chip measuring 10~mm $\times$ 4~mm. This chip can generate squeezed light in up to eight optical modes. To achieve this, the chip is \hbox{initialized} into four separate two-mode squeezed vacuum states. The squeezing happens between specific bichromatic mode pairs, each occupying one of the four spatially separated waveguide modes. An interferometer, consisting of a network of beam splitters and phase shifters, is used to effectively manipulate these spatial modes. This setup allows for a \hbox{user-programmable} gate sequence, corresponding to an SU(4) transformation applied to the spatial modes, synthesizing an eight-mode Gaussian state. This state is then measured on the Fock basis using eight separate photon-number-resolving detectors. For a visual representation of the machine's operation, refer to the equivalent quantum circuit diagram in Figure~\ref{fig8}b.

\subsection{Photon Loss}

Photon loss is a notable phenomenon in quantum optics that can impact quantum systems in various ways. When photons are lost or absorbed by the environment, it can result in quantum state degradation, such as Decoherence and State Dephasing, Reduced Fidelity, etc. However, this protocol demonstrates robustness~\cite{107}, functioning effectively even in the presence of photon loss, non-ideal sources, and imperfect detection. The protocol's ability to scale to handle a large number of photons provides strong evidence of its resilience, highlighting its practical application potential. This scalability is a significantly simpler task compared to building a universal quantum computer~\cite{108}.

\subsection{Security Considerations}

Experimental demonstrations of CV-QKD over metropolitan networks and field deployments highlight its potential for large-scale secure communication systems. Advancements in modulation techniques and protocol designs continue to bridge the gap between theoretical innovation and practical implementation, ensuring secure information transmission in the quantum era.

Despite these advancements, CV-QKD systems remain susceptible to specific vulnerabilities, such as quantum attacks and side-channel attacks. Quantum attacks, including collective Gaussian attacks, involve eavesdroppers performing collective measurements on transmitted quantum states to extract key information. Side-channel attacks exploit imperfections in system components, like the saturation attack on homodyne detectors, which can bias noise estimation and compromise security.

To enhance the robustness of CV-QKD systems, several mitigation strategies are proposed:
\begin{enumerate}
\item[A.] Enhanced Component Design: Improving the linearity and dynamic range of detectors reduces susceptibility to saturation attacks.

\item[B.] Advanced Security Protocols: Developing protocols that account for potential side-channel information leakage bolsters system resilience.

\item[C.] Regular System Audits: Conducting periodic security assessments helps identify and rectify vulnerabilities proactively.
\end{enumerate}

Incorporating machine learning (ML) and tensor networks into CV-QKD systems offers promising avenues for enhancing both functionality and security. ML techniques have been applied to various stages of CV-QKD protocols, including phase error estimation, excess noise estimation, state discrimination, parameter estimation, key sifting, and key rate estimation. These applications aim to alleviate implementation complexities and improve system performance~\cite{35}.

For instance, ML-assisted phase estimation can enhance carrier recovery processes, leading to more accurate phase compensation and reduced excess noise. This improvement is crucial for maintaining high secret key rates over long distances \cite{36,40}.

Moreover, tensor networks, which efficiently represent and manipulate high-dimensional data, can optimize the processing of quantum information in CV-QKD systems. Their application can lead to more efficient data handling and improved security analyses, particularly in complex quantum networks \cite{109,110,111}.

Measurement-device-independent quantum key distribution (MDI-QKD)~\cite{83} enhances security by eliminating all detector side-channel vulnerabilities \cite{112,113}, making it a robust solution for secure communications across various platforms. In fiber-based systems, MDI-QKD has achieved significant milestones, including successful key distribution over 404 km of ultralow-loss optical fiber, demonstrating its potential for long-distance applications~\cite{15}. For free-space channels, recent advancements have addressed challenges such as atmospheric turbulence. Notably, a study demonstrated MDI-QKD over a 19.2~km urban atmospheric link, marking a crucial step toward satellite-based quantum communications \cite{114,115}. In the realm of integrated photonics, the development of chip-based MDI-QKD systems has been pivotal. Researchers have successfully implemented an all-chip MDI-QKD system using silicon photonic technology, achieving a per-pulse key rate of 2.923 bits~\cite{116}. Furthermore, new methods can provide ease of use for MDI-QKD devices, for example by introducing plug-and-play capability~\cite{117}. These advancements collectively underscore the versatility and security of MDI-QKD across diverse communication infrastructures using different modulation methods, even with the presence of some imperfections such as the fast-fading channel \hbox{\cite{118,119,120,121,122,123,124,125,126}}.

\vspace*{-4pt}\enlargethispage{7pt}
\section{Conclusion}\label{sec6}

Quantum key distribution (QKD) protocols have significant theoretical advantages in terms of security against quantum computation algorithms like shore algorithms. However, there are still many challenges on the way. Some challenges like limited range, incompatibility with current networks, and other challenges that were mentioned in this paper are currently being addressed by ongoing research efforts with the aim of improving the speed, range, and scalability of QKD. Although solving or minimizing these problems seems feasible, there is still one inherent practical security issue that might be the actual Achilles heel of every QKD protocol. All classical methods to distribute a secure key are fundamentally insecure because in classical physics there is nothing preventing an eavesdropper from copying the key during its transit from Alice to Bob. This poses particular risk in the case of a Man-in-the-middle attack and necessitates the presence of a secure classical channel parallel to the quantum one~\cite{127}.

The proposed method of sending results of homodyne measurement over the classical channel is supposed to partially reduce the extent of this problem. This is because none of the keys could be directly reconstructed using the measurement results.

Modern CV-QKD research focuses on integrating these systems with existing optical communication infrastructure. The compatibility of CV-QKD with commercial telecom components, such as continuous-wave lasers and coherent receivers, has driven its rapid adoption. Furthermore, measurement-device-independent (MDI) and source-device-independent (SDI) schemes address practical security concerns, making CV-QKD increasingly viable for real-world applications.

Quantum networks, which leverage the principles of quantum entanglement and superposition, promise to revolutionize secure communication by enabling the transmission of quantum information across vast distances. A significant milestone in this field has been achieved with the successful entanglement of nanophotonic quantum memory \cite{128,129} nodes within a standard telecom network infrastructure.

In a groundbreaking study, researchers have demonstrated the entanglement of two nanophotonic quantum memory nodes separated by a physical distance and connected through standard telecom optical fibers~\cite{128}. Each node comprises an ensemble of rare-earth ions embedded in a nanophotonic cavity, functioning as a quantum memory capable of storing and retrieving entangled states. The entanglement process involves generating photon pairs, where one photon is stored in the quantum memory, and the other is transmitted through the telecom fiber to the distant node. By employing wavelength conversion techniques, the photon's wavelength is adapted to match the low-loss window of the telecom fiber, thereby minimizing transmission losses. Upon arrival at the second node, the photon interacts with the quantum memory, resulting in the entanglement of the two spatially separated nodes.

This achievement addresses several critical challenges in the realization of large-scale quantum networks. One is the Integration with Existing Telecom Infrastructure with the use of standard telecom fibers and wavelength conversion techniques ensures compatibility with current communication networks, facilitating the practical deployment of quantum networking technologies. Secondly, the nanophotonic design of the quantum memory nodes allows for compact and scalable integration, essential for building complex quantum networks with numerous interconnected nodes. At last, by effectively managing photon transmission losses through wavelength conversion and low-loss fibers, \hbox{entanglement} has been successfully distributed over significant distances, a crucial requirement for functional quantum networks.

The implications of this research are profound, paving the way for the development of secure quantum communication channels over existing telecom infrastructures. Future efforts will focus on enhancing the efficiency and fidelity of entanglement distribution, extending the achievable distances, and integrating more nodes to construct robust, large-scale quantum networks.

Additionally, Recent advances in coherent Ising machines have opened new avenues for generating robust\break entanglement in optical systems. Yanagimoto et~al.~\cite{130} demonstrated that by embedding entanglement generation within a measurement-feedback coherent Ising machine, it is possible to sustain quantum correlations among optical modes, thereby enhancing the overall performance of the system. In parallel, Inui and Yamamoto~\cite{131} reported that optically coupled coherent Ising machines not only produce significant entanglement but also exhibit nonclassical correlations such as quantum discord, even under realistic conditions. These findings suggest that integrating coherent Ising machine technology into quantum photonic platforms could provide scalable and resilient entanglement resources, complementing both continuous-variable and discrete-variable quantum communication protocols.

Consequently, Experimental efforts have successfully demonstrated CV-QKD over metropolitan networks and field deployments, showing its potential for large-scale secure communication systems. These advancements in modulation techniques and protocol designs continue to bridge the gap between theoretical innovation and practical implementation, ensuring the secure transmission of information in the quantum era.

\authorcontributions{Mobin Motaharifar and Mahmood Hasani contributed equally to this work, including conceptualization, writing, and formatting. Hasan Kaatuzian supervised the project. All authors have read and agreed to the published version of the manuscript.}

\funding{This research received no external funding.}

\conflictsofinterest{The authors declare no conflicts of interest.}

\dataavailability{Data is available upon request from corresponding authors.}


\begin{thebibliography}{000}

\bibitem{1} D. Stucki \textit{et~al.} (2011). ``Long-term performance of the SwissQuantum quantum key distribution network in a field environment''. \textit{New Journal of Physics}, 13: 12, 123001.

\bibitem{2} M. Sasaki \textit{et~al.} (2011). ``Field test of quantum key distribution in the Tokyo QKD Network''. \textit{Optics Express}, 19: 11, 10387--10409.

\bibitem{3} R. J. Hughes, J. E. Nordholt, K. P. McCabe, R. T. Newell, C. G. Peterson, and R. D. Somma (2013). ``Network-centric quantum communications with application to critical infrastructure protection''. \textit{arXiv preprint arXiv:1305.0305}.

\bibitem{4} H.-K. Lo, M. Curty, and K. Tamaki (2014). ``Secure quantum key distribution''. \textit{Nature Photonics}, 8: 8, \hbox{595--604}.

\bibitem{5} A. Orieux and E. Diamanti (2016). ``Recent advances on integrated quantum communications''. \textit{Journal of Optics}, 18: 8, 083002.

\bibitem{6} P. Sibson \textit{et~al.} (2017). ``Chip-based quantum key distribution''. \textit{Nature Communications}, 8: 1, 13984.

\bibitem{7} S. Tanzilli, A. Martin, F. Kaiser, M. P. De Micheli, O. Alibart, and D. B. Ostrowsky (2012). ``On the genesis and evolution of integrated quantum optics''. \textit{Laser \& Photonics Reviews}, 6: 1, 115--143.

\bibitem{8} P. Zhang \textit{et~al.} (2014). ``Reference-frame-independent quantum-key-distribution server with a telecom tether for an on-chip client''. \textit{Physical Review Letters}, 112: 13, 130501.

\bibitem{9} A. Politi, M. J. Cryan, J. G. Rarity, S. Yu, and J. L. O'Brien (2008). ``Silica-on-silicon waveguide quantum circuits''. \textit{Science}, 320: 5876, 646--649.

\bibitem{10} K. M. Davis, K. Miura, N. Sugimoto, and K. Hirao (1996). ``Writing waveguides in glass with a femtosecond laser''. \textit{Optics Letters}, 21: 21, 1729--1731.

\bibitem{11} J. G. Huang \textit{et~al.} (2017). ``Torsional frequency mixing and sensing in optomechanical resonators''. \textit{Applied Physics Letters}, 111: 11.

\bibitem{12} Y. Shi \textit{et~al.} (2018). ``Nanometer-precision linear sorting with synchronized optofluidic dual barriers''. \textit{Science Advances}, 4: 1, p. eaao0773.

\bibitem{13} Y. Z. Shi \textit{et~al.} (2018). ``Sculpting nanoparticle dynamics for single-bacteria-level screening and direct binding-efficiency measurement''. \textit{Nature Communications}, 9: 1, 815.

\bibitem{14} A. Boaron \textit{et~al.} (2018). ``Secure quantum key distribution over 421 km of optical fiber''. \textit{Physical Review Letters}, 121: 19, 190502.

\bibitem{15} H.-L. Yin \textit{et~al.} (2016). ``Measurement-device-independent quantum key distribution over a 404 km optical fiber''. \textit{Physical Review Letters}, 117: 19, 190501.

\bibitem{16} Y.-M. Li, X.-Y. Wang, Z.-L. Bai, W.-Y. Liu, S.-S. Yang, and K.-C. Peng (2017). ``Continuous variable quantum key distribution''. \textit{Chinese Physics B}, 26: 4, 040303.

\bibitem{17} C. H. Bennett and G. Brassard (2014). ``Quantum cryptography: Public key distribution and coin tossing''. \textit{Theoretical Computer Science}, 560: 7--11.

\bibitem{18} C. H. Bennett (1992). ``Quantum cryptography using any two nonorthogonal states''. \textit{Physical Review Letters}, 68: 21, 3121.

\bibitem{19} A. K. Ekert (1991). ``Quantum cryptography and Bell's theorem''. \textit{Quantum Measurements in Optics}, 413--418.

\bibitem{20} L. Ma \textit{et~al.} (2023). ``Practical continuous-variable quantum key distribution with feasible optimization parameters''. \textit{Science China Information Sciences,} 66: 8, 180507.

\bibitem{21} S. Sarmiento \textit{et~al.} (2022). ``Continuous-variable quantum key distribution over a 15 km multi-core fiber''. \textit{New Journal of Physics}, 24: 6, 063011.

\bibitem{22} Y. Zhang, Y. Bian, Z. Li, S. Yu, and H. Guo (2024). ``Continuous-variable quantum key distribution system: Past, present, and future''. \textit{Applied Physics Reviews}, 11: 1.

\bibitem{23} S. Pirandola (2021). ``Composable security for continuous variable quantum key distribution: Trust levels and practical key rates in wired and wireless networks''. \textit{Physical Review Research}, 3: 4, 043014.

\bibitem{24} N. Wang, S. Du, W. Liu, X. Wang, Y. Li, and K. Peng (2018). ``Long-distance continuous-variable quantum key distribution with entangled states''. \textit{Physical Review Applied}, 10: 6, 064028.

\bibitem{25} C. Weedbrook \textit{et~al.} (2012). ``Gaussian quantum information''. \textit{Reviews of Modern Physics}, 84: 2, 621--669.

\bibitem{26} G. Adesso, S. Ragy, and A. R. Lee (2014). ``Continuous variable quantum information: Gaussian states and beyond''. \textit{Open Systems \& Information Dynamics}, 21: 01n02, 1440001.

\bibitem{27} A. Ferraro, S. Olivares, and M. G. Paris (2005). ``Gaussian states in continuous variable quantum information''. \textit{arXiv preprint quant-ph/0503237}.

\bibitem{28} I. W. Primaatmaja, W. Y. Kon, and C. Lim (2024). ``Discrete-modulated continuous-variable quantum key distribution secure against general attacks''. \textit{arXiv preprint arXiv:2409.02630}.

\bibitem{29} N. Alshaer, T. Ismail, and H. Mahmoud (2024). ``Enhancing Performance of Continuous-Variable Quantum Key Distribution (CV-QKD) and Gaussian Modulation of Coherent States (GMCS) in Free-Space Channels under Individual Attacks with Phase-Sensitive Amplifier (PSA) and Homodyne Detection (HD)''. \textit{Sensors}, 24: 16, 5201.

\bibitem{30} S. B\"{a}uml, C. Pascual-Garc\'{\i}a, V. Wright, O. Fawzi, and A. Ac\'{\i}n (2024). ``Security of discrete-modulated continuous-variable quantum key distribution''. \textit{Quantum}, 8: 1418.

\bibitem{31} P. Papanastasiou, C. Ottaviani, and S. Pirandola (2021). ``Security of continuous-variable quantum key distribution against canonical attacks'', in \textit{2021 International Conference on Computer Communications and Networks (ICCCN)}: IEEE, 1--6.

\bibitem{32} A. G. Mountogiannakis, P. Papanastasiou, B. Braverman, and S. Pirandola (2022). ``Composably secure data processing for Gaussian-modulated continuous-variable quantum key distribution''. \textit{Physical Review Research}, 4: 1, 013099.

\bibitem{33} L. d. S. Aguiar, L. F. Borelli, J. A. Roversi, and A. Vidiella-Barranco (2022). ``Performance analysis of continuous-variable quantum key distribution using non-Gaussian states''. \textit{Quantum Information Processing,} 21: 8, 304.

\bibitem{34} N. Hosseinidehaj, A. M. Lance, T. Symul, N. Walk, and T. C. Ralph (2020). ``Finite-size effects in continuous-variable quantum key distribution with Gaussian postselection''. \textit{Physical Review A}, 101: 5, 052335.

\bibitem{35} N. K. Long, R. Malaney, and K. J. Grant (2022). ``A survey of machine learning assisted continuous-variable quantum key distribution''. \textit{Information}, 14: 10, 553.

\bibitem{36} H.-M. Chin, N. Jain, D. Zibar, U. L. Andersen, and T. Gehring (2021). ``Machine learning aided carrier recovery in continuous-variable quantum key distribution''. \textit{npj Quantum Information}, 7: 1, 20.

\bibitem{37} Q. Liao, J. Liu, A. Huang, L. Huang, Z. Fei, and X. Fu (2023). ``High-rate discretely-modulated continuous-variable quantum key distribution using quantum machine learning''. \textit{arXiv preprint arXiv:2308.03283}.

\bibitem{38} W.-B. Liu, C.-L. Li, Z.-P. Liu, M.-G. Zhou, H.-L. Yin, and Z.-B. Chen (2022). ``Theoretical development of discrete-modulated continuous-variable quantum key distribution''. (in English), \textit{Frontiers in Quantum Science and Technology}, Mini Review, 1: 985276, doi: 10.3389/frqst.2022.985276.

\bibitem{39} Y. Yan \textit{et~al.} (2023). ``Artificial key fingerprints for continuous-variable quantum key distribution''. \textit{Physical Review A}, 108: 1, 012601, doi: 10.1103/PhysRevA.108.012601.

\bibitem{40} C. Ding, S. Wang, Y. Wang, Z. Wu, J. Sun, and Y. Mao (2023). ``Machine-learning-based detection for quantum hacking attacks on continuous-variable quantum-key-distribution systems''. \textit{Physical Review A}, 107: 6, 062422, doi: 10.1103/PhysRevA.107.062422.

\bibitem{41} D. F. Walls (1983). ``Squeezed states of light''. \textit{Nature}, 306: 5939, 141--146.

\bibitem{42} H. Lin and Y. Shang (2024). ``Deterministic search on complete bipartite graphs by continuous time quantum walk''. \textit{arXiv preprint arXiv:2404.01640}.

\bibitem{43} B. Zhang and Q. Zhuang (2021). ``Entanglement formation in continuous-variable random quantum networks''. \textit{npj Quantum Information,} 7: 1, 33.

\bibitem{44} C. S. Hamilton, R. Kruse, L. Sansoni, S. Barkhofen, C. Silberhorn, and I. Jex (2017). ``Gaussian boson sampling''. \textit{Physical Review Letters}, 119: 17, 170501.

\bibitem{45} H.-S. Zhong \textit{et~al.} (2020). ``Quantum computational advantage using photons''. \textit{Science}, 370: 6523, 1460--1463.

\bibitem{46} H.-S. Zhong \textit{et~al.} (2021). ``Phase-programmable gaussian boson sampling using stimulated squeezed light''. \textit{Physical Review Letters}, 127: 18, 180502.

\bibitem{47} A. Czerwinski (2024). ``Quantum state tomography of photonic qubits with realistic coherent light sources''. \textit{Quantum Information \& Computation}, 24: 31--39.

\bibitem{48} R. Schnabel (2017). ``Squeezed states of light and their applications in laser interferometers''. \textit{Physics Reports}, 684: 1--51.

\bibitem{49} H. Vahlbruch \textit{et~al.} (2008). ``Observation of squeezed light with 10--dB quantum-noise reduction''. \textit{Physical Review Letters}, 100: 3, 033602.

\bibitem{50} H. Vahlbruch, M. Mehmet, K. Danzmann, and R. Schnabel (2016). ``Detection of 15 dB squeezed states of light and their application for the absolute calibration of photoelectric quantum efficiency''. \textit{Physical Review Letters}, 117: 11, 110801.

\bibitem{51} S. Du and Z. Bai (2024). ``State convertibility under genuinely incoherent operations''. \textit{Quantum Information \& Computation}, 24: 17--30.

\bibitem{52} S. Barzanjeh, S. Guha, C. Weedbrook, D. Vitali, J. H. Shapiro, and S. Pirandola (2015). ``Microwave quantum illumination''. \textit{Physical Review Letters}, 114: 8, 080503.

\bibitem{53} H. Wang \textit{et~al.} (2020). ``Observation of intensity squeezing in resonance fluorescence from a solid-state device''. \textit{Physical Review Letters}, 125: 15, 153601.

\bibitem{54} M. Sabatini, T. Bertapelle, P. Villoresi, G. Vallone, and M. Avesani (2024), ``Hybrid encoder for discrete and continuous variable QKD''. \textit{arXiv preprint arXiv:2408.17412}.

\bibitem{55} R. Loudon and P. L. Knight (1987). ``Squeezed light''. \textit{Journal of Modern Optics}, 34: 6--7, 709--759.

\bibitem{56} A. I. Lvovsky (2015). ``Squeezed light''. \textit{Photonics: Scientific Foundations, Technology and Applications}, 1: 121--163.

\bibitem{57} B. E. Saleh and M. C. Teich (2019). \textit{Fundamentals of Photonics}. John Wiley \& Sons.

\bibitem{58} D. Bouwmeester, J.-W. Pan, K. Mattle, M. Eibl, H. Weinfurter, and A. Zeilinger (1997). ``Experimental quantum teleportation''. \textit{Nature}, 390: 6660, 575--579.

\bibitem{59} M. B. Plenio and S. Virmani (2005). ``An introduction to entanglement measures''. \textit{arXiv preprint quant-ph/0504163}.

\bibitem{60} M. Motaharifar, H. Kaatuzian, and M. Hasani (2023). ``Possible teleportation of quantum states using squeezed sources and photonic integrated circuits'', in \textit{2023 5th Iranian International Conference on Microelectronics (IICM)}, IEEE, 227--232.

\bibitem{61} L. Albano, D. Mundarain, and J. Stephany (2002). ``On the squeezed number states and their phase space representations''. \textit{Journal of Optics B: Quantum and Semiclassical Optics}, 4: 5, 352.

\bibitem{62} M. A. Nielsen and I. L. Chuang (2010). \textit{Quantum Computation and Quantum Information}. Cambridge \hbox{University} Press.

\bibitem{63} M. Motaharifar and H. Kaatuzian (2023). ``Mach-Zehnder interferometer cell for realization of quantum computer; a feasibility study'', in \textit{2023 31st International Conference on Electrical Engineering (ICEE)}. IEEE, 762--767.

\bibitem{64} D. Dai and J. E. Bowers (2011). ``Novel ultra-short and ultra-broadband polarization beam splitter based on a bent directional coupler''. \textit{Optics Express}, 19: 19, 18614--18620.

\bibitem{65} J. Wang, D. Liang, Y. Tang, D. Dai, and J. E. Bowers (2013). ``Realization of an ultra-short silicon polarization beam splitter with an asymmetrical bent directional coupler''. \textit{Optics Letters}, 38: 1, 4--6.

\bibitem{66} T. C. Ralph (1999). ``Continuous variable quantum cryptography''. \textit{Physical Review A}, 61: 1, 010303.

\bibitem{67} M. Hillery (2000). ``Quantum cryptography with squeezed states''. \textit{Physical Review A}, 61: 2, 022309.

\bibitem{68} M. D. Reid (2000). ``Quantum cryptography with a predetermined key, using continuous-variable Einstein-Podolsky-Rosen correlations''. \textit{Physical Review A}, 62: 6, 062308.

\bibitem{69} N. J. Cerf, M. Levy, and G. Van Assche (2001). ``Quantum distribution of Gaussian keys using squeezed states''. \textit{Physical Review A}, 63: 5, 052311.

\bibitem{70} D. Gottesman and J. Preskill (2003). \textit{Quantum Information with Continuous Variables}. Springer Dordrecht.

\bibitem{71} F. Grosshans and P. Grangier (2002). ``Continuous variable quantum cryptography using coherent states''. \textit{Physical Review Letters}, 88: 5, 057902.

\bibitem{72} A. Leverrier and P. Grangier (2009). ``Unconditional security proof of long-distance continuous-variable quantum key distribution with discrete modulation''. \textit{Physical Review Letters}, 102: 18, 180504.

\bibitem{73} X. Su, W. Wang, Y. Wang, X. Jia, C. Xie, and K. Peng (2009). ``Continuous variable quantum key distribution based on optical entangled states without signal modulation''. \textit{Europhysics Letters}, 87: 2, 20005.

\bibitem{74} A. Denys, P. Brown, and A. Leverrier (2021). ``Explicit asymptotic secret key rate of continuous-variable quantum key distribution with an arbitrary modulation''. \textit{Quantum}, 5: 540.

\bibitem{75} L. S. Madsen, V. C. Usenko, M. Lassen, R. Filip, and U. L. Andersen (2012). ``Continuous variable quantum key distribution with modulated entangled states''. \textit{Nature Communications}, 3: 1, 1083.

\bibitem{76} V. C. Usenko and R. Filip (2011). ``Squeezed-state quantum key distribution upon imperfect reconciliation''. \textit{New Journal of Physics}, 13: 11, 113007.

\bibitem{77} C. Weedbrook, A. M. Lance, W. P. Bowen, T. Symul, T. C. Ralph, and P. K. Lam (2004). ``Quantum cryptography without switching''. \textit{Physical Review Letters,} 93: 17, 170504.

\bibitem{78} A. M. Lance, T. Symul, V. Sharma, C. Weedbrook, T. C. Ralph, and P. K. Lam (2005). ``No-switching quantum key distribution using broadband modulated coherent light''. \textit{Physical Review Letters}, 95: 18, 80503.

\bibitem{79} R. Garc\'{\i}a-Patr\'{o}n and N. J. Cerf (2009). ``Continuous-variable quantum key distribution protocols over noisy channels''. \textit{Physical Review Letters}, 102: 13, 130501.

\bibitem{80} J. Lodewyck \textit{et~al.} (2007). ``Quantum key distribution over 25 km with an all-fiber continuous-variable system''. \textit{Physical Review A---Atomic, Molecular, and Optical Physics}, 76: 4, 042305.

\bibitem{81} F. Grosshans, N. J. Cerf, J. Wenger, R. Tualle-Brouri, and P. Grangier (2003). ``Virtual entanglement and reconciliation protocols for quantum cryptography with continuous variables''. \textit{arXiv preprint quant-ph/0306141}.

\bibitem{82} G. Roberts \textit{et~al.} (2018). ``Patterning-effect mitigating intensity modulator for secure decoy-state quantum key distribution''. \textit{Optics Letters}, 43: 20, 5110--5113.

\bibitem{83} H.-K. Lo, M. Curty, and B. Qi (2012). ``Measurement-device-independent quantum key distribution''. \textit{Physical Review Letters}, 108: 13, 130503, doi: 10.1103/PhysRevLett.108.130503.

\bibitem{84} Y. Ding \textit{et~al.} (2017). ``High-dimensional quantum key distribution based on multicore fiber using silicon photonic integrated circuits''. \textit{npj Quantum Information}, 3: 1, 25.

\bibitem{85} C. Ma \textit{et~al.} (2016). ``Silicon photonic transmitter for polarization-encoded quantum key distribution''. \textit{Optica}, 3: 11, 1274--1278.

\bibitem{86} P. Sibson, J. E. Kennard, S. Stanisic, C. Erven, J. L. O'Brien, and M. G. Thompson (2017). ``Integrated silicon photonics for high-speed quantum key distribution''. \textit{Optica}, 4: 2, 172--177.

\bibitem{87} F. Najafi \textit{et~al.} (2015). ``On-chip detection of non-classical light by scalable integration of single-photon detectors''. \textit{Nature Communications}, 6: 1, 5873.

\bibitem{88} W. H. Pernice \textit{et~al.} (2012). ``High-speed and high-efficiency travelling wave single-photon detectors embedded in nanophotonic circuits''. \textit{Nature Communications}, 3: 1, 1325.

\bibitem{89} M. Ziebell \textit{et~al.} (2015). ``Towards on-chip continuous-variable quantum key distribution'', in \textit{The European Conference on Lasers and Electro-Optics}. Optica Publishing Group.

\bibitem{90} P. Jouguet, S. Kunz-Jacques, A. Leverrier, P. Grangier, and E. Diamanti (2013). ``Experimental demonstration of long-distance continuous-variable quantum key distribution''. \textit{Nature Photonics}, 7: 5, 378--381.

\bibitem{91} D. Huang, P. Huang, D. Lin, and G. Zeng (2016). ``Long-distance continuous-variable quantum key distribution by controlling excess noise''. \textit{Scientific Reports}, 6: 1, 19201.

\bibitem{92} F. Raffaelli \textit{et~al.} (2018). ``A homodyne detector integrated onto a photonic chip for measuring quantum states and generating random numbers''. \textit{Quantum Science and Technology}, 3: 2, 025003.

\bibitem{93} M. Rud\'{e} \textit{et~al.} (2018). ``Interferometric photodetection in silicon photonics for phase diffusion quantum entropy sources''. \textit{Optics Express}, 26: 24, 31957--31964.

\bibitem{94} F. Raffaelli, P. Sibson, J. E. Kennard, D. H. Mahler, M. G. Thompson, and J. C. Matthews (2018). ``Generation of random numbers by measuring phase fluctuations from a laser diode with a silicon-on-insulator chip''. \textit{Optics Express}, 26: 16, 19730--19741.

\bibitem{95} C. Abellan \textit{et~al.} (2016). ``Quantum entropy source on an InP photonic integrated circuit for random number generation''. \textit{Optica}, 3: 9, 989--994.

\bibitem{96} P. Kaur, A. Boes, G. Ren, T. G. Nguyen, G. Roelkens, and A. Mitchell, (2021). ``Hybrid and heterogeneous photonic integration''. \textit{APL Photonics}, 6,~6.

\bibitem{97} K. Jaksch \textit{et~al.} (2024). ``Composable free-space continuous-variable quantum key distribution using discrete modulation''. \textit{arXiv preprint arXiv:2410.12915}.

\bibitem{98} Y. Xu \textit{et~al.} (2021). ``Hybrid external-cavity lasers (ECL) using photonic wire bonds as coupling elements''. \textit{Scientific Reports}, 11: 1, 16426.

\bibitem{99} C. Koos \textit{et~al.} (2013). ``Photonic wire bonding: An enabling technology for heterogeneous multi-chip integration'', in \textit{Integrated Photonics Research, Silicon and Nanophotonics}. Optica Publishing Group.

\bibitem{100} M. Hasani, H. Kaatuzian, and M. Motaharifar, (2023). ``Stochastic mass-spring model for the generation of squeezed state of light'', in \textit{Laser Science}. Optica Publishing Group.

\bibitem{101} S. Zhao \textit{et~al.} (2024). ``Broadband amplitude squeezing at room temperature in electrically driven quantum dot lasers''. \textit{Physical Review Research}, 6: 3, L032021.

\bibitem{102} M. Hasani and M. Motaharifar, (2024). ``Experimental realization of spontaneous parametric down conversion''.

\bibitem{103} M. Rabiei, H. Kaatuzian, M. Hasani, and A. Shircharandabi, (2024). ``Analysis of squeezed light generation via SFWM in a Si$_{3}$N$_{4}$ microring resonator'', in \textit{Frontiers in Optics}. Optica Publishing Group.

\bibitem{104} A. Shircharandabi, H. Kaatuzian, M. Hasani, and M. Rabiei, (2024). ``Investigation of squeezed-state generation using SFWM in a SiO2 microring resonator'', in \textit{Laser Science}. Optica Publishing Group.

\bibitem{105} E. Engin \textit{et~al.} (2013). ``Photon pair generation in a silicon micro-ring resonator with reverse bias enhancement''. \textit{Optics Express}, 21: 23, 27826--27834.

\bibitem{106} J. M. Arrazola \textit{et~al.} (2021). ``Quantum circuits with many photons on a programmable nanophotonic chip''. \textit{Nature}, 591: 7848, 54--60.

\bibitem{107} M. A. Broome \textit{et~al.} (2013). ``Photonic boson sampling in a tunable circuit''. \textit{Science}, 339: 6121, 794--798.

\bibitem{108} A. Molina and J. Watrous, (2019). ``Revisiting the simulation of quantum Turing machines by quantum circuits''. \textit{Proceedings of the Royal Society A}, 475: 2226, 20180767.

\bibitem{109} R. Nagai, T. Tomono, and Y. Minato, (2021). ``Simulation of Continuous-Variable Quantum Systems with Tensor Network''. in \textit{2021 IEEE International Conference on Quantum Computing and Engineering (QCE)}. IEEE, 437--438.

\bibitem{110} Q. Zhang, H. Lai, J. Pieprzyk, and L. Pan, (2022). ``An improved quantum network communication model based on compressed tensor network states''. \textit{Quantum Information Processing}, 21: 7, 253.

\bibitem{111} Q. Zhang, H. Lai, and J. Pieprzyk, (2022). ``Quantum-key-expansion protocol based on number-state-entanglement-preserving tensor network with compression''. \textit{Physical Review A}, 105: 3, 032439.

\bibitem{112} A. Rubenok, J. A. Slater, P. Chan, I. Lucio-Martinez, and W. Tittel, (2013). ``Real-world two-photon interference and proof-of-principle quantum key distribution immune to detector attacks''. \textit{Physical Review Letters}, 111: 13, 130501, doi: 10.1103/PhysRevLett.111.130501.

\bibitem{113} T. Li, Z. Gao, and Z. Li, (2020). ``Measurement-device--independent quantum secure direct communication: \hbox{Direct} quantum communication with imperfect measurement device and untrusted operator''. \textit{Europhysics Letters}, 131: 6, 60001.

\bibitem{114} Y. Cao \textit{et~al.} (2020). ``Long-distance free-space measurement-device-independent quantum key distribution''. \textit{Physical Review Letters}, 125: 26, 260503.

\bibitem{115} Y.-H. Li \textit{et~al.} (2023). ``Free-space and fiber-integrated measurement-device-independent quantum key distribution under high background noise''. \textit{Physical Review Letters}, 131: 10, 100802.

\bibitem{116} L. Cao \textit{et~al.} (2020). ``Chip-based measurement-device-independent quantum key distribution using integrated silicon photonic systems''. \textit{Physical Review Applied}, 14: 1, 011001.

\bibitem{117} Q. Liao, Y. Wang, D. Huang, and Y. Guo, (2018). ``Dual-phase-modulated plug-and-play measurement-device-independent continuous-variable quantum key distribution''. \textit{Optics Express}, 26: 16, 19907--19920.

\bibitem{118} P. Wang, Y. Tian, and Y. Li, (2025). ``Advances in continuous variable measurement-device-independent quantum key distribution''. \textit{arXiv preprint arXiv:2502.16448}.

\bibitem{119} Y.-C. Zhang, Z. Li, S. Yu, W. Gu, X. Peng, and H. Guo, (2014). ``Continuous-variable measurement-device-independent quantum key distribution using squeezed states''. \textit{Physical Review A,}. 90: 5, 052325, doi: 10.1103/PhysRevA.90.052325.

\bibitem{120} L. Fan, Y. Bian, Y. Zhang, and S. Yu, (2022). ``Free-space continuous-variable quantum key distribution with imperfect detector against uniform fast-fading channels''. \textit{Symmetry}, 14: 6, 1271. https://www.mdpi.com/2073-8994/14/6/1271.

\bibitem{121} L. Ruppert \textit{et~al.} (2019). ``Fading channel estimation for free-space continuous-variable secure quantum communication''. \textit{New Journal of Physics}, 21: 12, 123036.

\bibitem{122} R. Zhao, J. Zhou, R. Shi, and J. Shi, (2024). ``Unidimensional continuous variable quantum key distribution under fast fading channel''. \textit{Annalen der Physik}, 536: 5, 2300401.

\bibitem{123} F. Yang, D. Qiu, and P. Mateus, (2023). ``Continuous-variable quantum secret sharing in fast-fluctuating channels''. \textit{IEEE Transactions on Quantum Engineering}, 4: 1--9.

\bibitem{124} P. Papanastasiou, C. Weedbrook, and S. Pirandola, (2018). ``Continuous-variable quantum key distribution in uniform fast-fading channels''. \textit{Physical Review A}, 97:  3, 032311, doi: 10.1103/PhysRevA.97.032311.

\bibitem{125} S. Pirandola \textit{et~al.} (2015). ``MDI-QKD: Continuous-versus discrete-variables at metropolitan distances''. \textit{arXiv preprint arXiv:1506.06748}.

\bibitem{126} P. Wang, X. Wang, and Y. Li, (2019). ``Continuous-variable measurement-device-independent quantum key distribution using modulated squeezed states and optical amplifiers''. \textit{Physical Review A}, 99: 4, 042309, doi: 10.1103/PhysRevA.99.042309.

\bibitem{127} E. Diamanti, H.-K. Lo, B. Qi, and Z. Yuan, (2016). ``Practical challenges in quantum key distribution''. \textit{npj Quantum Information}, 2: 1, 1--12.

\bibitem{128} C. M. Knaut \textit{et~al.} (2024). ``Entanglement of nanophotonic quantum memory nodes in a telecom network''. \textit{Nature}, 629: 8012, 573--578.

\bibitem{129} \'{E}. Dumur \textit{et~al.} (2021). ``Quantum communication with itinerant surface acoustic wave phonons''. \textit{npj Quantum Information}, 7: 1, 173.

\bibitem{130} R. Yanagimoto, P. L. McMahon, E. Ng, T. Onodera, and H. Mabuchi, (2019). ``Embedding entanglement generation within a measurement-feedback coherent Ising machine''. \textit{arXiv preprint arXiv:1906.04902}.

\bibitem{131} Y. Inui and Y. Yamamoto, (2020). ``Entanglement and quantum discord in optically coupled coherent Ising machines''. \textit{Physical Review A,} 102: 6, 062419.

\end{thebibliography}
\end{document}